\input{aipcheck}

\documentclass[
    ,final            
  ]
  {aipproc}

\layoutstyle{6x9}

\def\gtrsim{\mathrel{\hbox{\rlap{\hbox{\lower4pt\hbox{$\sim$}}}\hbox{$>$}}}}
\def\lesssim{\mathrel{\hbox{\rlap{\hbox{\lower4pt\hbox{$\sim$}}}\hbox{$<$}}}}
\newcommand{\kms}{km s$^{-1}$}
\newcommand{\hi}{H{\sc i }}

\newcommand{\msun}{M$_{\odot}$}
\newcommand{\cmsq}{cm$^{-2}$}

\begin{document}

\title{The \hi distribution in the outskirts of M33 with the ALFALFA survey}



\author{M. Grossi}{
  address={INAF-Osservatorio Astrofisico di Arcetri, Florence, Italy}
}

\author{C. Giovanardi}{
  altaddress={INAF-Osservatorio Astrofisico di Arcetri, Florence, Italy}
}

\author{E. Corbelli}{
  altaddress={INAF-Osservatorio Astrofisico di Arcetri, Florence, Italy} 
}

\begin{abstract}
 Spiral galaxies appear to be dynamical systems whose disks are still forming at the current epoch and which continue to accrete mass. The presence of extraplanar gas in spirals indicates that galactic halos can contain at least part of the material needed to fuel the star formation activity in  their disks. Here we present the analysis of the ALFALFA survey data in the region of  M33 aimed at searching high velocity clouds around this Local Group galaxy.
We find a varied population of \hi clouds  with masses ranging between  $4 \times 10^4$ and $\sim 10^6$ M$_{\odot}$.
We also detect an \hi complex at anomalous velocity, whose extragalactic nature cannot be firmly established. We estimate that the total amount of neutral hydrogen mass associated to these clouds is $\lesssim 10^7$ M$_{\odot}$.
\end{abstract}

\maketitle


\section{Introduction}
Recent 21-cm observations of nearby spiral galaxies indicate that their halos can contain up
to about 15$\%$ of the total amount of neutral hydrogen (Oosterloo, Fraternali \& Sancisi 2007 and references therein).
In the Local Group both the Milky Way and Andromeda show a population of high velocity clouds distributed in their halos, whose nature and origin is still uncertain (Wakker \& van Woerden, Braun \& Burton 2000, Putman et al. 2002, Thilker et al. 2004, Westmeier et al. 2005).
Contrary to these two galaxies M33 is fairly isolated, and its recent evolution  seems to have been rather smooth without showing evidence for  recent mergers. However several observational facts suggest that the disk of M33  is still in the process of accreting matter.  A diffuse and low column density
\hi filament has been detected to the north west of M33 connecting this galaxy to M31 (Braun \& Thilker 2004), one \hi cloud has been recently identified  around M33 (Westmeier et al. 2005).
To inspect the \hi content of the halo of M33 and search for additional \hi clouds we have used the 21-cm observations provided by the Arecibo Legacy Fast ALFA (ALFALFA) survey (Giovanelli et al. 2005) and here we present the preliminary results of this analysis.

\begin{figure}
  \includegraphics[width=0.7\textwidth]{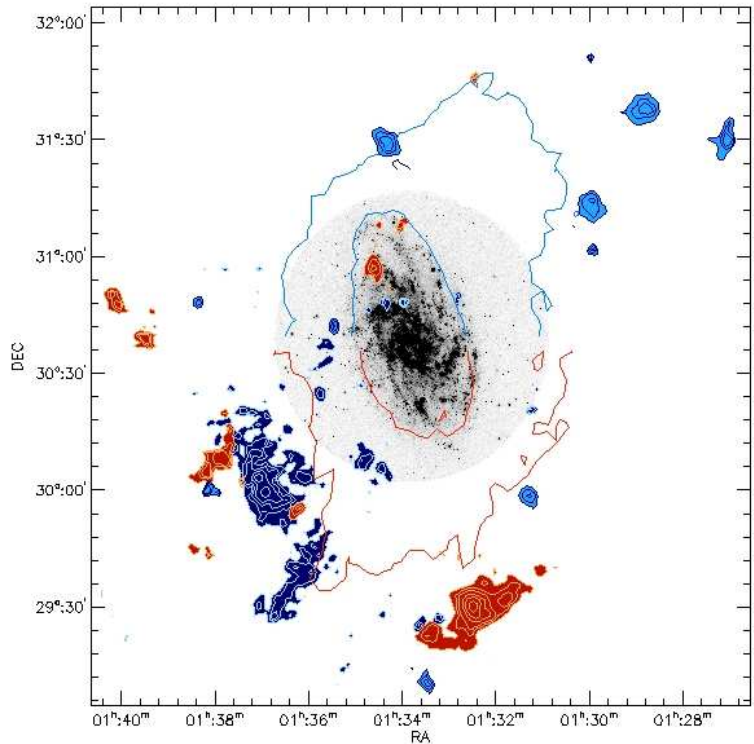}
  \caption{The spatial distribution of \hi clouds around M33. Contours indicate \hi column densities between $3 \times 10^{18}$ cm$^{-2}$ and $4 \times 10^{19}$ cm$^{-2}$. The colors indicate the different radial velocity of the clouds with respect to the systemic of M33 (-180 \kms):  red ($-80$ \kms $< V < -180$ \kms) and light blue ($-180$ \kms $< V < -320$ \kms) correspond to the clouds with velocities comparable to the rotation of the disk. In dark blue we show the clouds with an anomalous kinematic ($-320$ \kms $< V < -400$ \kms). The contours overlaid on the GALEX image of M33 indicate the \hi disk at  $5\times 10^{19}$ \cmsq and $10^{21}$ \cmsq. The blue and red colors show the approaching and  receding sides of the disk.}
\end{figure}

\section{Overview of the results}

The ALFALFA data
reveal a population of high velocity \hi clouds  and more extended \hi
complexes around M33. We can distinguish two types of detections at 21-cm: i) clouds that have radial velocities comparable to the rotation of the M33 disk (-80 \kms $< V <$ -320 \kms; ii)
  diffuse  \hi complexes (R $>$ 1 kpc) with anomalous radial velocities (-320 \kms $< V <$ -400 \kms (Figure 1).
 The first type of clouds are more likely to be associated to M33.
 Their masses range between $2 \times 10^4$ and $4 \times 10^6$ \msun and their average column densities are below $10^{19}$ \cmsq . From Figure 1 one can see that they appear to be distributed along one main axis which is oriented towards Andromeda (to the north west of M33).
The more massive ones (M$_{HI} > 2\times 10^5$ \msun) have an elongated or filament-like shape, and in some cases appear to connect to the disk at a certain velocity, as if they are in the process of being accreted to the disk or tidal stripped.
The more compact objects instead are overall isolated in space from the disk of M33 and less massive. If we assume that the clouds are close to the internal dynamical equilibrium and they are self-gravitating, their virial masses are between 100 and 2000 times larger than the \hi mass.

At smaller radial velocities ($V < -320$ \kms) we have resolved a \hi complex that had been serendipitously discovered by Thilker et al. (2002) with Westerbork observations of M33. The coarser spatial resolution of their observations (10 times ALFALFA) prevented to firmly establish the connection with M33. The complex is  at about one degree from the centre of M33 ($\sim$ 15 kpc).
Two main substructures can be distinguished from the ALFALFA data (Figure 2) showing a clear velocity gradient with the northern one being closer in velocity to the disk ($-365 < V < -395$ km s$^{-1}$). At $V \sim -370$  km s$^{-1}$ and ($\alpha$, $\delta$) = (01$^h$:35$^m$:30$^s$,30$^{\circ}$:30${^{\prime}}$) there appear to be faint clumps of gas pointing towards the center of M33, which may give indication for a possible interaction with the galaxy (see  Figure 2).  We cannot exclude that this is a cloud not gravitationally bound to M33 and higher sensitivity observations of the region  are needed to establish this hypothesis.
However given the larger extension of this complex with respect to the other clouds and the anomalous velocity, it is likely that this is a local \hi complex.

\begin{figure}
  \includegraphics[width=0.6\textwidth]{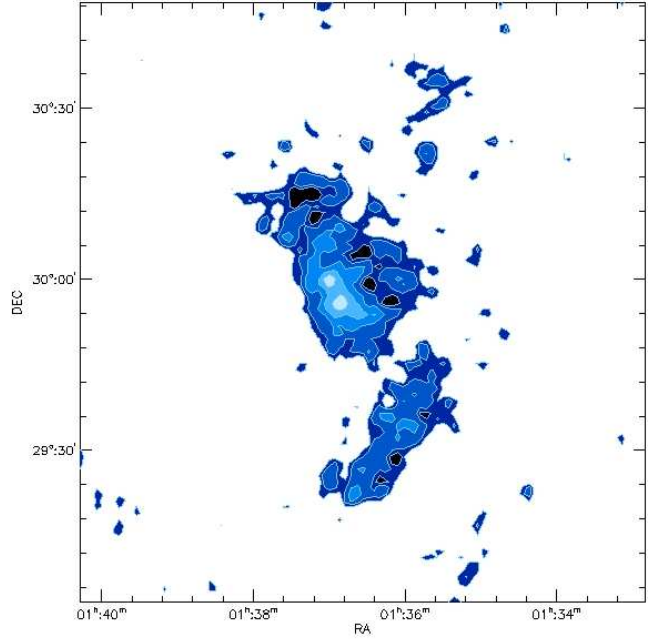}
  \caption{The column density map of the \hi complex with anomalous velocity to the south east of M33. Contours are drawn from 0.3 to 1.2 $\times 10^{19}$ \cmsq.}
\end{figure}

\end{document}